\makeatletter
\@ifundefined{@parse@version@dash}{%
\def\@parse@version#1{\@parse@version@0#1}
\def\@parse@version@#1/#2/#3#4#5\@nil{%
\@parse@version@dash#1-#2-#3#4\@nil}
\def\@parse@version@dash#1-#2-#3#4#5\@nil{%
  \if\relax#2\relax\else#1\fi#2#3#4 }
}{}
\makeatother

\documentclass[%
 reprint,
 amsmath,amssymb,
 aps,
]{revtex4-2}

\usepackage{float}
\usepackage[colorlinks]{hyperref}

\usepackage{graphicx}
\usepackage{dcolumn}
\usepackage{bm}

\usepackage{color}

\hypersetup{
	setpagesize=false,
	bookmarksnumbered=true,%
	bookmarksopen=true,%
	colorlinks=true,
	linkcolor=blue,
	citecolor=blue,
	urlcolor=black,
}

\begin{document}

\title{
Lu/Se Substitution Effect on Magnetic Properties \\ of Yb-Based Zigzag Chain Semiconductor YbCuS$_{2}$
}%

\author{Fumiya Hori$^1$}
\email{hori.fumiya.36s@st.kyoto-u.ac.jp}
\author{Hiroyasu Matsudaira$^1$}
\author{Shunsaku Kitagawa$^1$}
\author{Kenji Ishida$^1$}
\email{kishida@scphys.kyoto-u.ac.jp}
\affiliation{$^1$Department of Physics, Kyoto University, Kyoto 606-8502, Japan}

\author{Souichiro Mizutani$^2$}
\author{Hirotaka Shirai$^2$}
\author{Takahiro Onimaru$^2$}
\affiliation{$^2$\mbox{Department of Quantum Matter, Graduate School of Advanced Science and Engineering, Hiroshima University,} \\ Higashihiroshima 739-8530, Japan}

\begin{abstract}
\section*{Abstract}
{
We have investigated the changes in the magnetic properties on YbCuS$_2$ by Lu or Se substitutions from a microscopic perspective.
In general, it is expected that  nonmagnetic Lu substitution dilutes the magnetic Yb$^{3+}$ concentration,
and that Se substitution induces the negative pressure in YbCuS$_2$.
The $^{63/65}$Cu-nuclear quadrupole resonance (NQR) measurements on polycrystalline Yb$_{0.9}$Lu$_{0.1}$CuS$_{2}$ and YbCu(S$_{0.9}$Se$_{0.1}$)$_{2}$ revealed that the Se substitution leads to larger lattice expansion compared to the Lu substitution.
The antiferromagnetic transition temperature $T_{\rm N}$ decreased from 0.95~K in the unsubstituted system to 0.75~K in both substituted systems.
Furthermore, the $T$-linear behavior of the $^{63}$Cu-NQR spin-lattice relaxation rate $1/T_1$, suggesting the presence of the gapless quasiparticle excitations, was observed even in the substituted systems, and the value of $1/T_1T$ increases.
These results under the chemical substitutions are opposite to those under pressure as previously reported, and are consistent with the expectation.
Our systematic study indicates a clear relationship between the magnetic ordered states and the quasiparticle excitations in YbCuS$_2$.
}

\end{abstract}

\maketitle

\section{Introduction}

Frustrated magnets have been intensively studied for more than 50 years, owing to their potential to give rise to novel nonmagnetic ground states; quantum spin liquid~\cite{frustrated-magnets, QSL-review, dmit-131_thermal_conductivity, Herbertsmithite-neutron, Yb2Ti2O7_thermal_conductivity, Kitaev, Kitaev-review, RuCl3-kasahara}, valence bond solid~\cite{Majumdar, CuGeO3-1, CuGeO3-2, SS_Momoi,  SrCu2⁢(BO3)⁢2, Hall_effect_triplons}, and spin nematic states~\cite{spin_nematic_triangle, spin_nematic_triangle-2, Nakatsuji_Ni⁢Ga2⁢S4, zigzag2, FM-AFM-zigzag, LiCuVO4_Magnetization}.
Furthermore, such systems may also induce unconventional charge-neutral quasiparticle excitations, including spinons~\cite{frustrated-magnets, QSL-review, dmit-131_thermal_conductivity, Herbertsmithite-neutron}, magnetic monopoles~\cite{frustrated-magnets, M_monopoles_spin_ice, Yb2Ti2O7_thermal_conductivity}, Majorana fermions~\cite{Kitaev, Kitaev-review, QSL-review, RuCl3-kasahara}, visons~\cite{Kitaev, Kitaev-review, frustrated-magnets} and triplons~\cite{SS_Momoi, SrCu2⁢(BO3)⁢2, Hall_effect_triplons}.
In recent years, there has been a growing focus on magnetic frustration in insulating or semiconducting compounds based on rare earth ions from Ce ($4f^1$) to Yb($4f^{13}$)~\cite{Yb_frustration, YbMgGaO4-1, YbMgGaO4-2, YbMgGaO4-3, NaYbSe2, NaYbSe2_spinon_fermi_surface, NaYbO2, NaYbS2, Yb_triangular_1, Yb_triangular_2, Yb_triangular_3, Ohmagari1, Ohmagari2, Hori2022, Hori2023, RAgSe2, Hori_YbAgSe2, Saito1, Saito2, Saito3}.
The interplay between strong spin–orbit coupling and crystalline electric field of rare-earth-based magnetism leads to the anisotropic exchange interactions.
These result in the emergence of non-trivial magnetic states not observed in standard quantum magnets based on 3$d$ transition metals~\cite{Yb_frustration}.
For instance, Yb-based insulators YbMgGaO$_4$~\cite{YbMgGaO4-1, YbMgGaO4-2, YbMgGaO4-3} and NaYb$Ch_2$ ($Ch =$ O, S, Se)~\cite{NaYbSe2, NaYbSe2_spinon_fermi_surface, NaYbO2, NaYbS2}, with Yb$^{3+}$ triangular lattices, have been proposed to exhibit exotic ground states, such as $Z_2$ spin liquids and spinon Fermi surfaces~\cite{ Yb_triangular_1, Yb_triangular_2, Yb_triangular_3}.

Chemical substitution is widely used in the study of quantum spin systems, particularly for probing the effects of magnetic frustrations and interaction modulations.
In some frustrated systems, 
substitution-induced long-range magnetic ordered states have been observed despite its absence in the unsubstituted systems~\cite{CuGeO3_substitution, Li2ScMo3O8_substitution, Rb2Cu3SnF12_substitution, BiCu2PO6_substitution}.
In the case of the above-mentioned Yb-based triangular-lattice magnets, the effects of the nonmagnetic Lu$^{3+}$ substitution, which dilutes the Yb$^{3+}$ ions, have been studied in detail~\cite{NaYbSe2_Lu_substitution, NaYbSe2_Lu_substitution_thermal, NaYbO2_Lu_substitution, NaYbS2_Lu_substitution}.
Interestingly, despite the Lu substitution, these systems do not exhibit a long-range magnetic order.
Instead, intriguing phenomena such as the substitution-induced Yb-Yb dimer states~\cite{NaYbSe2_Lu_substitution, NaYbO2_Lu_substitution} and the impurity-robust gapped excitations~\cite{NaYbSe2_Lu_substitution_thermal} have been proposed.

\begin{figure}[t]
\centering
\includegraphics[width=8cm]{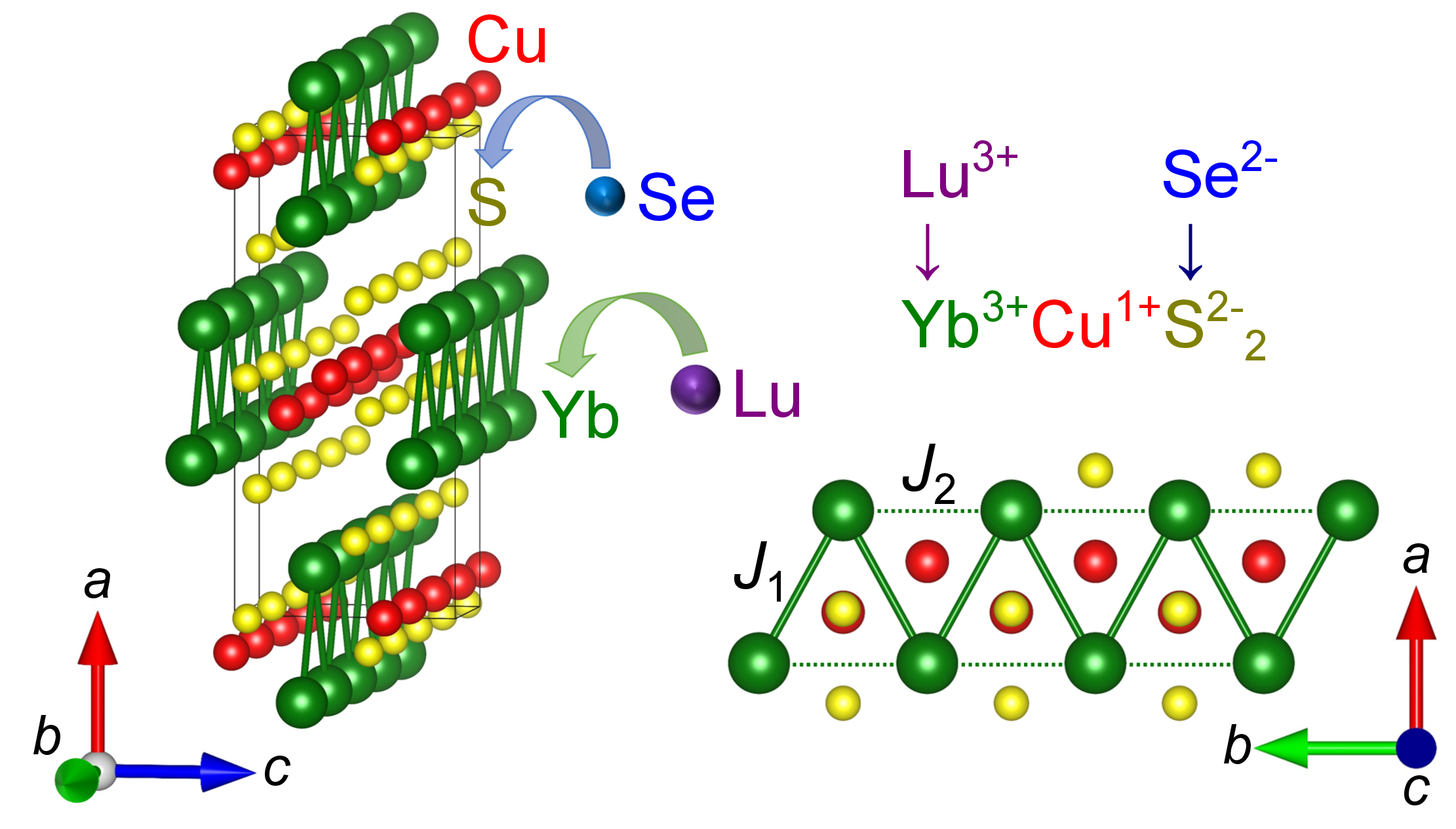} 
\caption{(a) Crystal structure of YbCuS$_2$ with the space group $Pnma$ (No. 62, $D_{2h}^{16}$)~\cite{YbCuS2_Pnma_structure};
Yb$^{3+}$ or S$^{2-}$ sites were partially substituted with Lu$^{3+}$ or Se$^{2-}$, respectively.
The magnetic frustration due to the competition between nearest-neighbor exchange interactions $J_1$ and next-nearest-neighbor exchange interactions $J_2$ is expected.
The solid line represents the unit cell.
The structural image was depicted using the VESTA program~\cite{vesta}.
}
\label{structure}
\end{figure}

A Yb-based magnetic semiconductor YbCuS$_2$ has attracted much attention as one of the rare-earth-based frustrated magnets.
This compound has Yb$^{3+}$ zigzag chains as shown in Fig.~\ref{structure}, where the frustration effect driven by the competition between nearest-neighbor exchange interactions $J_1$ and next-nearest-neighbor exchange interactions $J_2$ is expected.
Several unique properties, resulting from this frustration effect have been consistently observed~\cite{Ohmagari1, Ohmagari2, Hori2022, Hori2023}.
YbCuS$_2$ shows a first-order magnetic transition at $T_{\rm N} \sim 0.95$~K, characterized by the strong divergence of the specific heat and the coexistence of the paramagnetic (PM) and antiferromagnetic (AFM) states.
Below $T_{\rm N}$, an incommensurate AFM structure is confirmed by the $^{63/65}$Cu-nuclear quadrupole resonance (NQR) measurements~\cite{Hori2023} and the  neutron diffraction experiments~\cite{Onimaru_YbCuS2_neutron}.
The ordered moment is much smaller than that anticipated from the crystalline electric field ground state, reflecting the presence of the frustration effect derived from the Yb$^{3+}$ zigzag chains.
Below 0.5~K, the $T$-linear behavior of $1/T_1$ is observed.
This $T$-linear behavior cannot be explained by conventional magnon excitations and indicates the emergence of the gapless charge-neutral quasiparticle excitations in YbCuS$_2$~\cite{Hori2023}.

Furthermore, we have investigated the pressure effect on YbCuS$_2$ in the previous study~\cite{Hori_NQR_pressure}. 
At 1.6 GPa, $T_{\rm N}$ increases to 1.17 K, and the magnetic structure changes to a commensurate one, which can be regarded as an odd-parity magnetic multipole order~\cite{Hayami_JPSJ_review}. 
Remarkably, pressure suppresses the gapless quasiparticle excitations.
While these results indicate that YbCuS$_2$ hosts an intriguing  quantum state sensitive to the changes in the exchange interactions, the origin of these properties remains unclear.
It is possible that these characteristics may be explained by the recent theoretical model of the one-dimensional (1D) frustrated zigzag chains with the anisotropic exchange interactions~\cite{Saito1, Saito2, Saito3}. 
Therefore, further experimental investigations on YbCuS$_2$, focusing on tuning the exchange interaction parameters, are crucial.

Recently, substituted systems Yb$_{1-x}$Lu$_x$CuS$_{2}$ ($x < 0.3$) and YbCu(S$_{1-y}$Se$_y$)$_2$ ($y < 0.1$) were synthesized, in which Yb$^{3+}$ or S$^{2-}$ sites were partially substituted with Lu$^{3+}$ or Se$^{2-}$, respectively~\cite{YbCuS2_substitution}.
While the nonmagnetic Lu substitution dilutes the magnetic Yb$^{3+}$ concentration,
the Se substitution expands lattice constants with larger ionic radius than S.
From the powder X-ray diffraction measurements, the Lu substitution results in less than a 0.1\% change in the lattice constants for $x < 0.3$.
In contrast, Se-10\% substitution causes approximately a 0.4\% expansion along the $a$, $b$, and $c$ axes, suggesting significant lattice distortion.
However, in both cases, the magnetic susceptibilities follow the Curie-Weiss law, with no significant changes in the Weiss temperatures or the effective magnetic moments.
The specific-heat measurements suggest that the transition temperature $T_{\rm N}$ decreases by both Lu and Se substitutions.

In this study, we have performed $^{63/65}$Cu-NQR measurements on polycrystalline Yb$_{0.9}$Lu$_{0.1}$CuS$_{2}$ (Lu-10\%) and YbCu(S$_{0.9}$Se$_{0.1}$)$_{2}$ (Se-10\%) to investigate the changes in the magnetic properties by the chemical substitutions from a microscopic point of view.
The NQR peak position $^{63}\nu_{\rm Q}$ shifts with the Se substitution, while the Lu substitution does not affect the $^{63}\nu_{\rm Q}$ value so much.
This indicates the presence of the larger lattice expansion in Se-10\% than that in Lu-10\%.
It is confirmed that the AFM transition temperature decreases from $T_{\rm N} \sim 0.95$~K for YbCuS$_2$ to $\sim 0.75$~K for both Lu- and Se-substituted systems.
The $T$-linear behavior, suggesting the presence of the gapless quasiparticle excitation, was observed even in the substituted systems with an increase in $1/T_1T$.
These behaviors of $T_{\rm N}$ and $1/T_1T$ under the chemical substitutions contrast with those against pressure, showing a systematic change.
These results indicates a clear relationship between the magnetic ordered states and the quasiparticle excitations in
YbCuS$_2$.

\section{Experimental}

Polycrystalline samples of Lu-10\% and Se-10\% were synthesized by the melt-growth method.
A conventional spin-echo technique was used for the  $^{63/65}$Cu-NQR measurements.
The $^{63}$Cu (natural abundance: 69.1\%) and $^{65}$Cu (natural abundance: 30.9\%) nuclei have spin $I = 3/2$ (gyromagnetic ratios of $^{63}\gamma/2\pi = 11.289$~MHz/T and $^{65}\gamma/2\pi = 12.093$~MHz/T, and quadrupole moments of $^{63}Q = -0.220 \times 10^{-24}$~cm$^{2}$ and $^{65}Q = -0.204 \times 10^{-24}$~cm$^{2}$, respectively).
A $^3$He-$^4$He dilution refrigerator was used for the $^{63/65}$Cu-NQR measurements down to 0.1~K.
The peak position $^{63}\nu_{\rm Q}$ and the full width at half maximum (FWHM) of the PM $^{63}$Cu-NQR signal in Se-10\% were estimated by fitting the data with two Lorentzian curves as shown in the dotted curves of Fig.~\ref{spectrum}(a) since the $^{63}$Cu- and $^{65}$Cu-NQR signals overlap and a broad spectrum was observed.
The $^{63}$Cu-$1/T_1$ was evaluated by fitting the relaxation curve of the nuclear magnetization after the saturation to a theoretical function for the nuclear spin $I = 3/2$.
$1/T_1$ can be determined by a single relaxation component down to $T_{\rm N}$. 
However, short component was observed in the relaxation curve below $T_\mathrm{N}$ due to the appearance of the internal fields. Thus, we picked up the slowest components as described in previous study~\cite{Hori2023}.

\section{Results and Discussion}

\begin{table*}[t]
  \caption{The peak position $^{63}\nu_{\rm Q}$, the FWHM of the paramagnetic $^{63}$Cu-NQR signal, the AFM transition temperature $T_{\rm N}$, and the critical exponents of AFM transition $\beta$ in YbCuS$_2$, Lu-10\% and Se-10\%.}
  \centering
  \begin{tabular}{cccccc}
\hline \hline
Sample & pressure (GPa) & $^{63}\nu_{\rm Q}$ (MHz) & FWHM (MHz) & $T_{\rm N}$ (K) &  $\beta$\\ \hline
YbCuS$_2$ & 0 & 9.29 & 0.02 & 0.95 & 0.05 \\
 & 0.9 & 9.13 & 0.04  & 1.07 & 0.09  \\
 & 1.6 & 9.01 & 0.05  & 1.17 & 0.12  \\  
Lu-10\% & 0 & 9.29 & 0.13 & 0.75 & 0.36  \\
Se-10\% & 0 & 9.44 & 1.01  & 0.75 & 0.28 \\ \hline \hline
    \end{tabular}
    \label{table1}
\end{table*}

\begin{figure*}[t!]
\centering
\includegraphics[width=16cm]{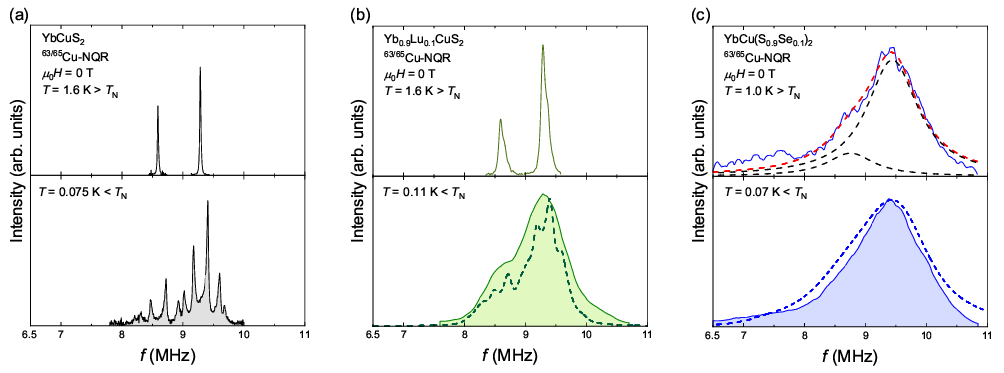} 
\caption{(a) The $^{63/65}$Cu-NQR spectra of the unsubstituted YbCuS$_2$ at 1.6~K (upper panel) and 0.075~K (lower panel).
(b) The $^{63/65}$Cu-NQR spectra of Yb$_{0.9}$Lu$_{0.1}$CuS$_2$ at 1.6~K (upper panel) and 0.11~K (lower panel); the dashed curve represents the convolution of the AFM $^{63/65}$Cu-NQR spectrum in the unsubstituted YbCuS$_2$ and the PM $^{63}$Cu-NQR signal in Lu-10\%.
(c) The $^{63/65}$Cu-NQR spectra of YbCu(S$_{0.9}$Se$_{0.1}$)$_2$ at 1.0~K (upper panel) and 0.07~K (lower panel); the dashed curves of the upper panel show fitting of two Lorentzians corresponding to $^{63}$Cu and $^{65}$Cu signals.
The dashed curve of the lower panel denotes the convolution of the AFM $^{63/65}$Cu-NQR spectrum in the unsubstituted YbCuS$_2$ and the PM $^{63}$Cu-NQR signal in Se-10\%.
}
\label{spectrum}
\end{figure*}

The upper panel of Fig.~\ref{spectrum} shows the $^{63/65}$Cu-NQR spectra above $T_{\rm N}$ in the unsubstituted YbCuS$_2$, Lu-10\%, and Se-10\%.
The peak position $^{63}\nu_{\rm Q}$ and FWHM of the PM $^{63}$Cu-NQR signal are listed in Table~\ref{table1}.
The spectra exhibit noticeable broadening in the substituted systems, which can be attributed to the distribution of the electric field gradient (EFG) at around Cu nuclei caused by the randomness effect due to the chemical substitutions.
It is noteworthy that the Se substitution makes the NQR spectrum extraordinarily broader than the Lu substitution.
This suggests that the EFG at the Cu site is determined mainly with the S configuration, and that the randomness is introduces in the inter-chain interactions.
Comparing the peak position $^{63}\nu_{\rm Q}$, it shifts from 9.29~MHz in the unsubstituted system to 9.44~MHz in Se-10\%, while the Lu substitution results in no significant changes.
According to the results of the powder X-ray diffraction~\cite{YbCuS2_substitution}, the Lu substitution leads to minimal changes in the lattice constants, whereas the Se substitution causes significant expansion.
The increase of $^{63}\nu_{\rm Q}$ in Se-10\% can be attributed to the presence of the lattice expansion, leading to a large variation in the EFG.
Note that pressure does not significantly affect FWHM in YbCuS$_2$, as presented in Table~\ref{table1}.
This suggests that pressure is a cleaner method for tuning parameters than the chemical substitutions.
Furthermore, pressure decreases the value of $\nu_{\rm Q}$, whereas the Se substitution increases it. This implies that the Se substitution corresponds to a negative pressure effect, consistent with the increase in lattice constants.


Below $T_{\rm N}$,  multiple peaks are observed in the unsubstituted sample as shown in the lower panel of  Fig.~\ref{spectrum}(a).
In contrast, the Lu-10\% and Se-10\% substituted samples exhibit a single broader signal below $T_{\rm N}$. 
The broadening of these signals could arise from either the distribution of $\nu_{\rm Q}$ due to the randomness of the substitution effect, or changes in the internal magnetic fields due to the altered magnetic structure below $T_{\rm N}$.
To estimate the effect of the distribution of $\nu_{\rm Q}$, we calculated the convolution given by
\begin{flalign}
(I_{\rm AFM} * L_{\rm PM})(f) = \sum_{f'} I_{\rm AFM}(f') \, L_{\rm PM}(f - f').
\end{flalign}
Here, $f$ and $f'$ represent the frequencies, and $I_{\rm AFM}(f)$ denotes the intensity of the AFM spectrum in the unsubstituted system.
$L_{\rm PM}(f)$ is the Lorentzian function with the same FWHM as the $^{63}$Cu-NQR PM signal in Lu-10\% or Se-10\%, corresponding to the distribution of $\nu_{\rm Q}$.
Since $^{63}\nu_{\rm Q}$ is different in each system, the curves obtained after the convolution calculation are shifted to the differences in $^{63}\nu_{\rm Q}$. 
The AFM spectra in both Lu-10\% and Se-10\% can be partially explained by the convolution curves as indicated by the dotted curves in the lower pannel of Fig.~\ref{spectrum}.
Therefore, one origin of the AFM spectrum change is the distribution of $\nu_{\rm Q}$ due to the randomness of the substitution effect. 
However, it is difficult from the current NQR spectra to determine whether or not the magnitude of the internal magnetic fields is changed by the chemical substitution.
Thus, to determine the size of the ordered moments and the magnetic structures, other measurements such as the neutron scattering are necessary.

\begin{figure}[t!]
\centering
\includegraphics[width=8cm]{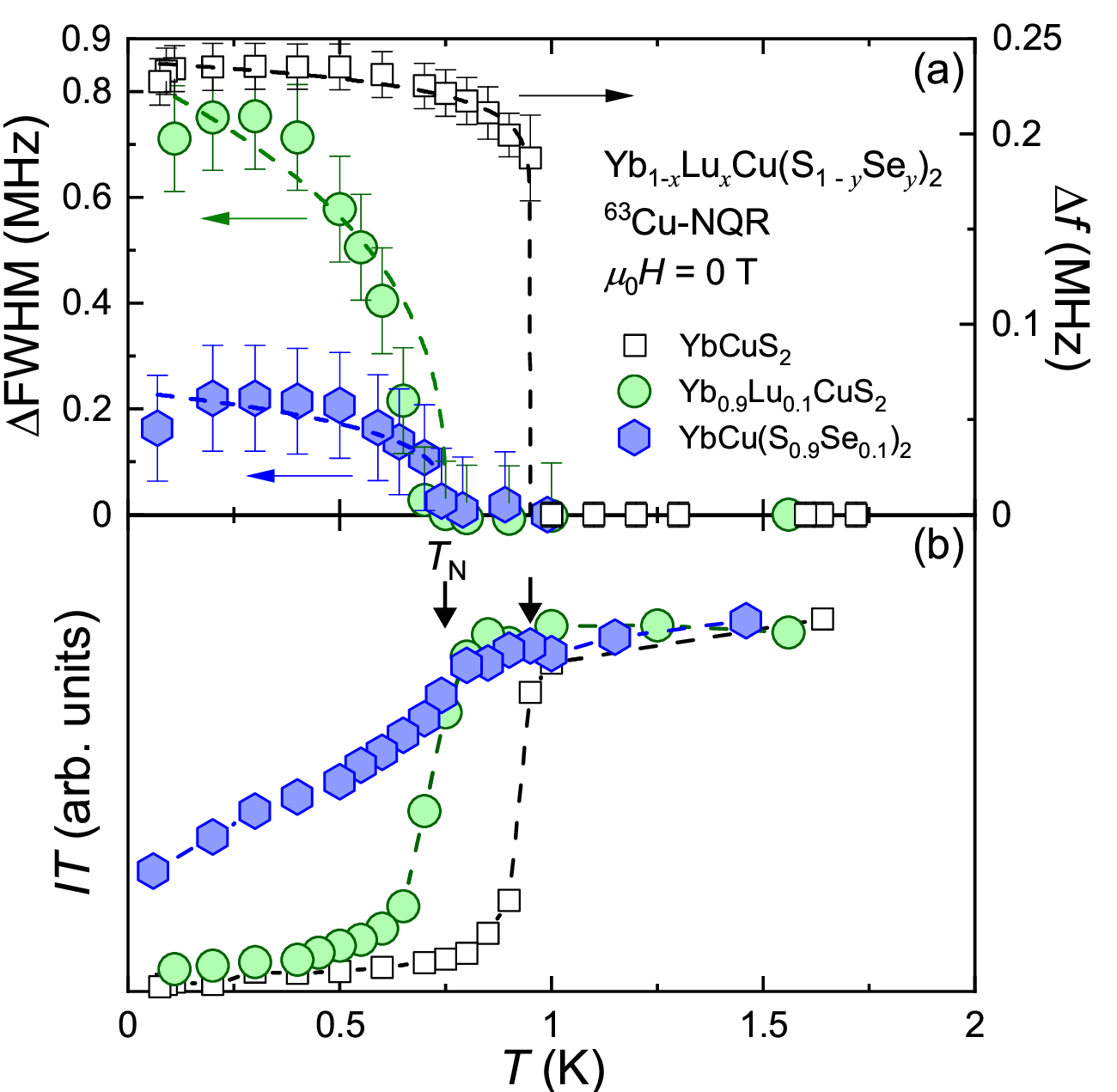} 
\caption{(a) Temperature dependence of change of full width
at half maximum $\Delta$FWHM from PM state (left axis) on the substituted systems and frequency difference of splitting peaks $\Delta f$ (right axis) on the unsubstituted system; the squares, circles, and hexagons denote $\Delta f$ in the unsubstituted YbCuS$_2$, $\Delta$FWHM in Yb$_{0.9}$Lu$_{0.1}$CuS$_2$ and YbCu(S$_{0.9}$Se$_{0.1}$)$_2$, respectively.
The dashed curves represent a fit with Eq.~(\ref{beta}).
(b) Temperature dependence of the products of the $^{63}$Cu-NQR peak intensity and the temperature $I(T)T$; the squares, circles, and hexagons denote $I(T)T$ in the unsubstituted YbCuS$_2$, Yb$_{0.9}$Lu$_{0.1}$CuS$_2$ and YbCu(S$_{0.9}$Se$_{0.1}$)$_2$, respectively.}
\label{IT}
\end{figure}

To track the substitution effect on the transition temperature $T_{\rm N}$, we investigated the temperature dependence of the change in linewidth $\Delta$FWHM, the frequency difference of splitting peaks $\Delta f$ and the NQR peak intensity $I (T)$, as shown in Fig.~\ref{IT}.
Here,  
$\Delta f$ was estimated from two peaks with the highest intensity in the unsubstituted systems, and  $I(T)$ was measured at the frequencies of the PM NQR peak.
In the PM state, the products of the NQR peak intensity and temperature $I(T)T$ are almost constant against temperature.
However, deviation from $I(T) T = \text{const.}$ occurs due to signal splitting or broadening of the spectrum.
We define $T_{\rm N}$ as the temperature below which $\Delta$FWHM and $\Delta f$ increase and $I(T)T$ decreases rapidly. 
As previously reported~\cite{Hori2023}, it was confirmed that the unsubstituted system exhibits an AFM order at $T_{\rm N} \sim 0.95$~K. 
In the substituted systems, both Lu-10\% and Se-10\%, 
$T_{\rm N}$ decreases to 0.75~K, which is consistent with the specific-heat results~\cite{YbCuS2_substitution}.
Note that the gradual change in $I(T)T$ below $T_{\rm N}$ in Se-10\% is due to the broad spectrum above $T_{\rm N}$, which makes $T_{\rm N}$ distributed.

As reported in the previous study~\cite{Hori2023}, the PM and AFM signals coexist in a certain temperature region on the unsubstituted systems, suggesting the presence of a first-order phase transition.
However, in the substituted systems, the linewidth is significantly broader even above $T_{\rm N}$, and it increases gradually below $T_{\rm N}$, making it difficult to determine whether PM and AFM signals coexist.
Here, to investigate how the first-order feature of phase transition in YbCuS$_2$  is affected by the substitutions, we estimated the critical exponent $\beta$ of the order parameter.
We assume that the internal magnetic fields $H_{\rm int}$ follow
\begin{flalign}
H_\mathrm{int}(T) = H_\mathrm{int}(0)\left[\frac{T_{\rm N} - T}{T_{\rm N}}\right]^{\beta},
\label{beta}
\end{flalign}
and $\Delta f$ on the unsubstituted system are proportional to $H_{\rm int}$
as shown in the dashed curve of Fig.~\ref{IT}(a).
Since it is difficult to estimate $\Delta f$ in the substituted systems due to the broad spectrum, we estimated $\beta$ by assuming that $\Delta$FWHM is proportional to $H_{\rm int}$.
The estimated $\beta$ are listed in Table~\ref{table1}.
For the unsubstituted system, $\beta \sim 0.05$. This small value of $\beta$ indicates the characteristics of the first-order phase transition.
On the other hand, in Lu-10\% and Se-10\%, 
$\beta \sim 0.36$ and $\beta \sim 0.28$, respectively.
The value of $\beta$ increases by the chemical substitutions and become close to the conventional mean-field value ($\beta = 0.5$).
This result suggests that the chemical substitutions weaken the first-order characteristics of the phase transition.

\begin{figure}[t]
\centering
\includegraphics[width=8cm]{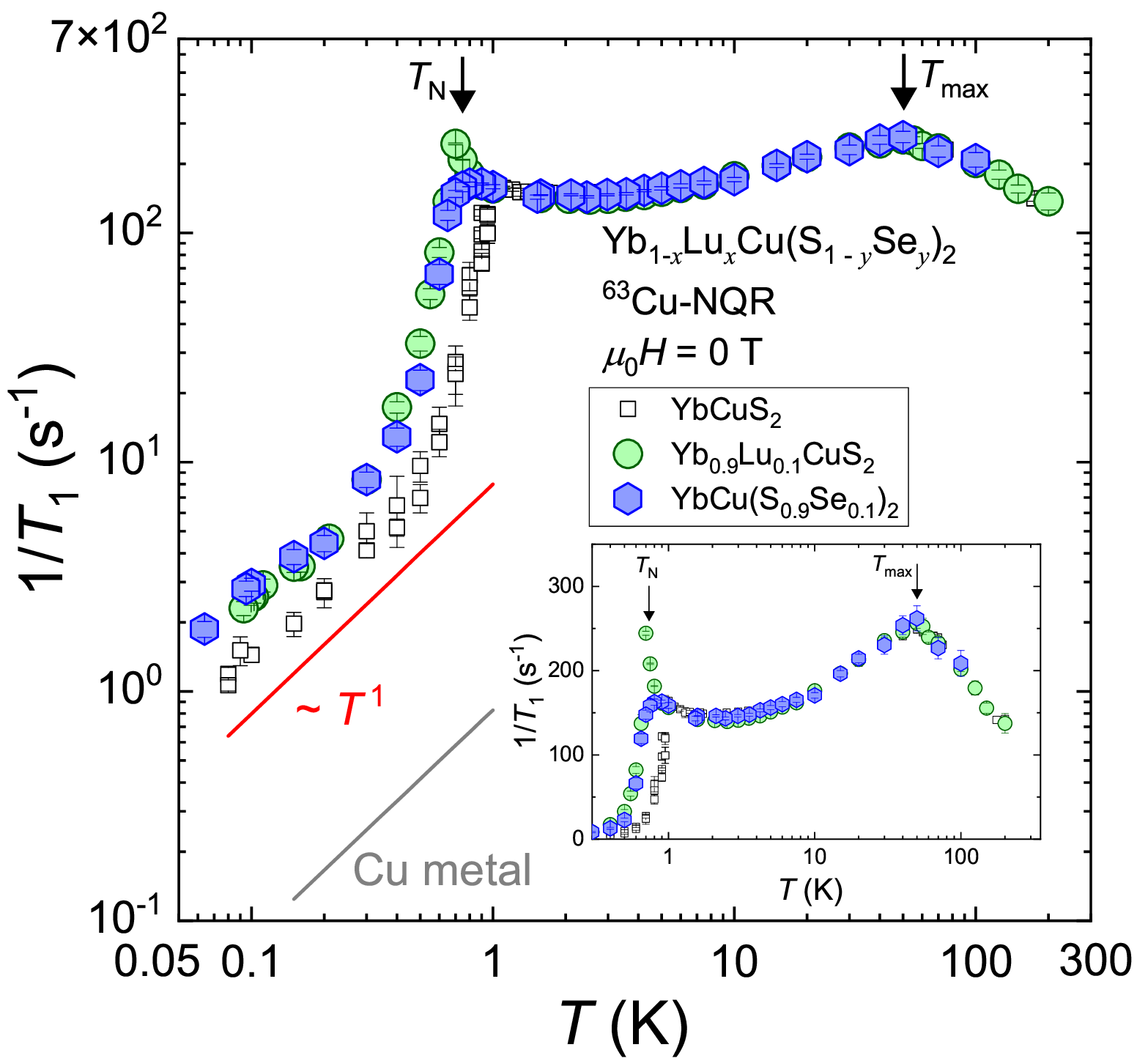} 
\caption{Temperature dependence of $1/T_1$ on Yb$_{1-x}$Lu$_x$Cu(S$_{1-y}$Se$_y$)$_2$; the squares, circles, and hexagons denote $1/T_1$ in the unsubstituted YbCuS$_2$, Yb$_{0.9}$Lu$_{0.1}$CuS$_2$ and YbCu(S$_{0.9}$Se$_{0.1}$)$_2$, respectively.
The inset also shows the temperature dependence of $1/T_1$.}
\label{1_T1}
\end{figure}

Figure~\ref{1_T1} shows the temperature dependence of the $^{63}$Cu-NQR $1/T_1$.
Despite the introduction of the chemical substitutions, no significant change in the anomaly at $T_{\rm max} \sim 50$~K was observed. 
This anomaly is also unchanged under pressure~\cite{Hori_NQR_pressure}.
The anomaly at $T_{\rm max}$ is robust against the substitutions and pressure.
The changes in $T_{\rm N}$ due to the substitutions are also confirmed from the temperature dependence of $1/T_1$.
In the Lu-substituted system, $1/T_1$ shows a sharp anomaly at $T_{\rm N}$, whereas in the Se-substituted system, it exhibits a more gradual change.
This may suggest the distribution in $T_{\rm N}$ by the Se substitution.
Interestingly, even with substitution applied, the gapless behavior with $1/T_1 \sim T$ is observed similarly to the unsubstituted system.
This suggests that this gapless quasiparticle excitation is robust against the chemical substitutions.
The value of $1/T_1T$ at 0.1~K increases from $\sim14$~s$^{-1}$ K$^{-1}$ in the unsubstituted system to $\sim25$~s$^{-1}$ K$^{-1}$ in Lu-10\% and $\sim29$~s$^{-1}$ K$^{-1}$ in Se-10\%.
The cause of the change in $1/T_1T$ will be discussed later.

\begin{figure}[t]
\centering
\includegraphics[width=8cm]{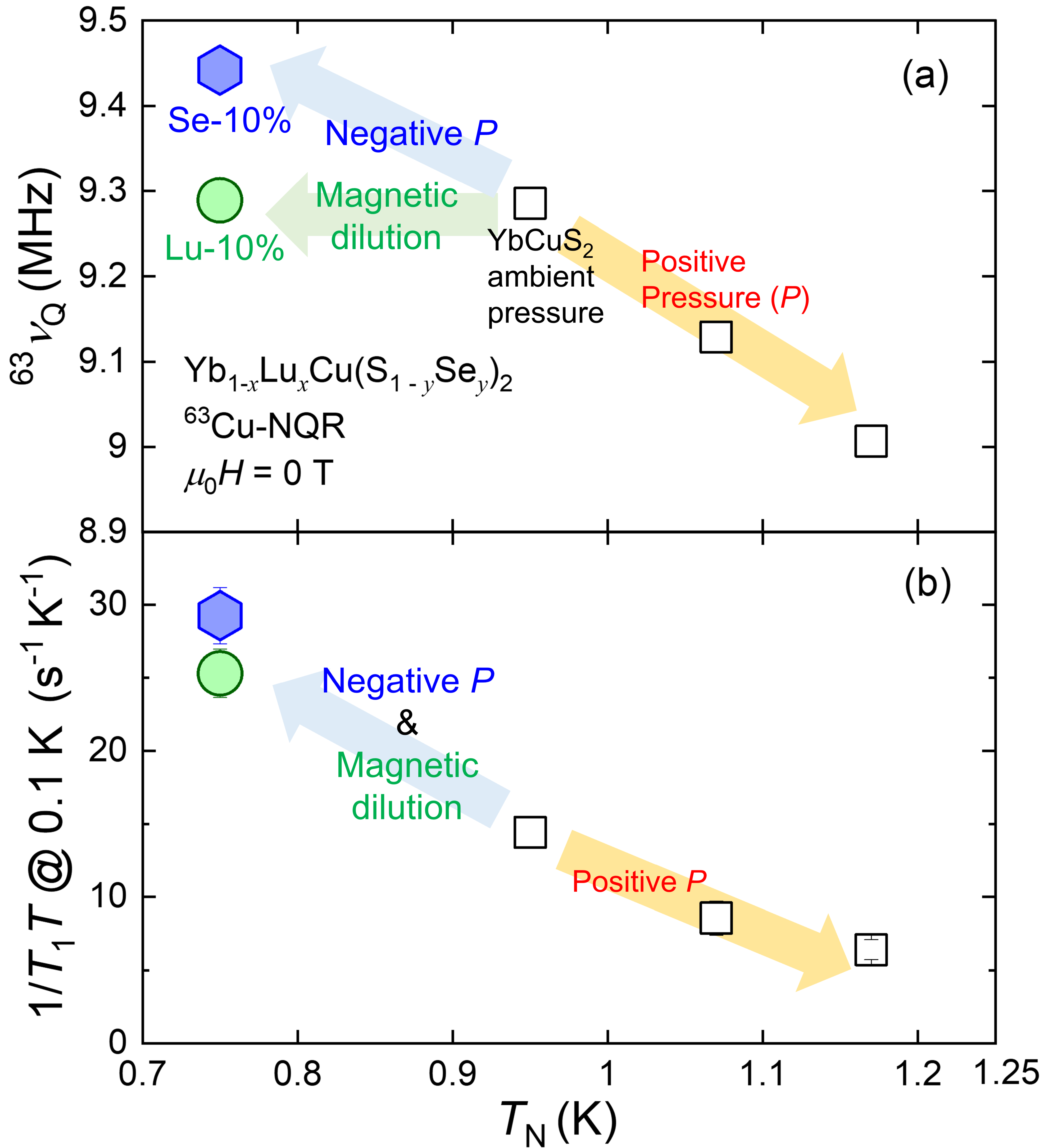} 
\caption{
(a) Relationship between $^{63}\nu_{\rm Q}$ and $T_{\rm N}$ in Yb$_{1-x}$Lu$_x$Cu(S$_{1-y}$Se$_y$)$_2$.
(b) Relationship between $T_{\rm N}$ and the value of $1/T_1T$ at 0.1~K.}
\label{phase_diagram}
\end{figure}

Here, we discuss the difference in the roles of Lu and Se substitutions on YbCuS$_2$.
Figure~\ref{phase_diagram}(a) shows the relationship between $^{63}\nu_{\rm Q}$ and $T_{\rm N}$.
As noted above, $^{63}\nu_{\rm Q}$ increase with the significant broadening by the Se substitution, while $^{63}\nu_{\rm Q}$ decrease without such the broadening by applying pressure.
Thus, the Se substitution corresponds to the negative pressure and randomness effect.
On the other hand, the Lu substitutions induce no significant change in $^{63}\nu_{\rm Q}$ and simply dilutes magnetic Yb$^{3+}$ ions.
These substitutions seem to produce different effects in principle.
However, both Lu-10\% and Se-10\% exhibit the same transition temperature, $T_{\rm N} \sim 0.75$~K.
In the theoretical 1D zigzag chain model, a long-range magnetic order cannot occur at $J_1 \sim J_2$, which is relevant to YbCuS$_2$~\cite{Saito1, Saito2, Saito3}.
Therefore, the discrepancy between the ideal 1D model and the experimental results is naturally attributed to the inter-chain coupling $J_3$.
The Se substitution expands the cell volume
and introduces the randomness in the coupling between the 1D chains.
These result in the reduction and the inhomogeneity in $J_{3}$, leading to a decrease and distribution in $T_{\rm N}$.
In contrast, the Lu substitution does not affect the cell volume as mentioned above.
Theoretically, $T_{\rm N}$ of the diluted quasi-1D antiferromagnets shows a simple reduction as a function of doping concentration of nonmagnetic ions due to the spin dilution effect~\cite{TN_doped_1D}, 
which is confirmed experimentally in various 1D compounds~\cite{Sr2CuO3_Pd_substitution}.
Thus, both Se and Lu substitutions reduce $T_{\rm N}$, which results in nearly the same $T_{\rm N}$ coincidentally.

Next, we discuss the relationship between the gapless quasiparticle excitations and the exhange interactions in YbCuS$_2$.
The chemical substitutions increase $1/T_1T$, while pressure decreases $1/T_1T$, as shown in Fig.~\ref{phase_diagram}(b).
Two possible mechanisms may explain the changes in $1/T_1T$.
First, the change in the intra-chain interactions $J_1$ and $J_2$ may influence the gapless excitations.
Since pressure decreases the $J_2/J_1$ ratio~\cite{Hori_NQR_pressure}, it is possible that the chemical substitutions have the opposite effect.
Second, the decrease in $T_{\rm N}$ by applying substitutions may increase fluctuations, leading to an enhance of gapless excitations. 
In typical quantum critical phenomena~\cite{QCP_Sachdev, QCP_Coleman, QCP_heavy-fermion}, decreasing $T_{\rm N}$ to destabilize a ordered state makes the system close to the quantum critical point, thereby enhancing fluctuations. 
Since both Lu-10\% and Se-10\% exhibit almost the same $T_{\rm N}$, the similar $1/T_1T$ values observed in these systems can be understood.
Note that, in either scenario, our results indicate that the origin of the gapless excitations is related to the intra-chain interactions, which contribute to quantum fluctuations.
The Lu and Se substitutions may not reduce magnetic frustration but instead weaken the magnetic order potentially leading to the enhancement of the gapless quasiparticle excitations.
To confirm these scenarios, further systematic studies with varying substitution concentrations or higher pressure are required, which is now in progress.

Finally, we discuss the origin of the gapless excitations in YbCuS$_2$.
In the previous report~\cite{Hori2023}, we mentioned that the origin of the gapless quasiparticle excitations in YbCuS$_2$ may be attributed to phasons in incommensurate ordered states~\cite{(TMTSF)2PF6_SeNMR, NMR_SDW} or spinons in low-dimensional systems~\cite{volborthite1, Zn-brochantite}.
However, such gapless excitations are typically suppressed and become gapped in the presence of impurities~\cite{Sr2CuO3_substitution_thermal, Sr2CuO3_substitution_NMR, SrCuO2_substitution_NMR, spin_chain_substitution_neutron, phason_impurity},
which seems unlikely in the case of YbCuS$_2$.
Recently, it has been theoretically proposed that the zigzag chains with the anisotropic exchange interactions, originating from the Yb$^{3+}$ ions, may give rise to gapless emergent quasiparticles called ``nematic particles"~\cite{Saito1, Saito2, Saito3}. 
While it is possible that the nematic particle is the origin of the gapless excitiation in YbCuS$_2$, the impurity effect for the nematic particle remains unclear.
In any case, our experimental results provide a constraint on the theoretical model that these excitations are robust against to the chemical substitutions.

\section{Conclusion} 

In conclusion, we have investigated the changes of the magnetic properties in YbCuS$_2$ by the Lu or Se substitutions from a microscopic point of view.
The NQR peak position $^{63}\nu_{\rm Q}$ shifts by the Se substitution, while the Lu substitution results in no significant change, indicating the presence of larger lattice expansion in Se-10\% than in Lu-10\%.
Both chemical substitutions lead to a decrease in the AFM transition temperature $T_{\rm N}$.
Despite these changes, the anomaly in $1/T_1$ at $T_{\rm max}$ remains unchanged in both the Lu- and Se-substituted systems.
Furthermore, the $T$-linear behavior of $1/T_1$, suggesting the presence of the gapless quasiparticle excitation, was observed even in the substituted systems with the increase of the value of $1/T_1T$.
The changes in $T_{\rm N}$ and $1/T_1T$ under the chemical substitutions are opposite to those under pressure, and a systematic change is observed.
These results highlight a relationship between the origin of the magnetic ordered states and quasiparticle excitations in YbCuS$_2$.

\begin{acknowledgments}
The authors would like to thank S. Yonezawa, H. Saito, and C. Hotta for valuable discussions. 
This work was supported by Grants-in-Aid for Scientific Research (KAKENHI Grant Nos. JP20KK0061, JP20H00130, JP21K18600, JP22H04933, JP22H01168, JP23H01124, JP23K22439, JP23K25821, JP23H04866, JP23H04870, JP23KJ1247 and JP24K00574) from the Japan Society for the Promotion of Science, by JST SPRING (Grant No. JPMJSP2110) from Japan Science and Technology Agency, by research support funding from the Kyoto University Foundation, by ISHIZUE 2024 of Kyoto University Research Development Program, and by Murata Science and Education Foundation.
In addition, liquid helium is supplied by the Low Temperature and Materials Sciences Division, Agency for Health, Safety and Environment, Kyoto University.
\end{acknowledgments}

\end{document}